\documentclass[prb,amsmath,amssymb,a4paper]{revtex4}

\usepackage{verbatim}
\usepackage{graphicx}
\usepackage[utf8]{inputenc}
\usepackage{hyperref}
\usepackage{caption}
\usepackage{epstopdf}
\usepackage{listings}

\newcommand{\be}{\begin{equation}}
\newcommand{\ee}{\end{equation}}

\newcommand{\md}{\mathrm{d}}
\newcommand{\nn}{\nonumber}

\begin{document}

\title{Nanotechnological Structure for Observation of Current Induced Contact Potential Difference and Creation of Effective Cooper Pair Mass-Spectroscopy}
\author{Todor M. Mishonov}
\email{mishonov@bgphysics.eu}
\affiliation{Institute of Solid State Physics, Bulgarian Academy of Sciences, 72 Tzarigradsko Chaussee Blvd., BG-1784 Sofia, Bulgaria}
\affiliation{Physics Faculty, St. Clement of Ohrid University of Sofia, 5 James Bourchier Blvd., BG-1164 Sofia}

\author{Albert Varonov}
\email{varonov@issp.bas.bg}
\affiliation{Institute of Solid State Physics, Bulgarian Academy of Sciences, 72 Tzarigradsko Chaussee Blvd., BG-1784 Sofia, Bulgaria}

\date{10 March, 12:51, 2020}

\begin{abstract}
Changes of the electron work-function of a superconductor proportional to the square of the current density $\Delta \phi = -\beta j^2$ are known as Bernoulli effect in superconductors or current induced Contact Potential Difference (CPD). 
The temperature dependent constant $\beta(T;m^{\star})$ is parametrized by the effective mass of Cooper pairs $m^{\star}$. 
In such a way the study of the Bernoulli effect leads to creation of Cooper pair mass-spectroscopy. 
In this paper a short review on the Bernoulli effect in superconductors is given
and a proposed experimental set-up for its measurement is described in detail.
The experiment requires standard electronic equipment and can be implemented in every laboratory related to physics of superconductivity.
This experimental set-up for observation of current induced CPD requires nano-technological hybrid superconductor structures with atomically clean interfaces and measurement of nano-volt signals with capacitive coupling to the sample.
\end{abstract}

\maketitle

\section{Introduction. $e^*$ and $m^*$. 
A historical perspective}
Starting from the formula for velocity 
$\mathbf{v}=(\hbar\nabla\theta-e^*\mathbf{A})/m^*$
for a charged super-fluid with wave-function 
$\Psi=\sqrt{n}\,\mathrm{e}^{\mathrm{i}\theta}$
F.~London\cite{London:35} derived the well-known relation between the 
variation of the vector-potential $\delta\mathbf{A}$ and the current density 
$\mathbf{j}=e^*n\mathbf{v}.$
This local approximation 
$\delta\mathbf{j}(\mathbf{r})
=-(e^{*2}n/m^*)\,\delta\mathbf{A}(\mathbf{r})$ 
leads to the well-known formulae for the London penetration depth
$1/\lambda^2=e^{*2}n/\varepsilon_0c^2 m^*$
and the quantization of magnetic flux\cite{London:50} with flux quantum 
$\Phi_0=2\pi\hbar [c]/e^*.$
Hereafter the notation $[c]$ is only in Gaussian system, in SI $[c]$ should be substituted by $1$.
In the relatively exotic case of rotating superconductor with local velocity of the lattice
$\mathbf{V}$ to the potential momentum $m_0\mathbf{V}$ a term containing the
inertial mass of the free charge carriers $m_0$ is added and the local expression for the current reads 
$\mathbf{j}
=e^*n(\hbar\nabla\theta-e^*\mathbf{A}
-m_0\mathbf{V})/m^*.$
This complete result gives the possibility for a precise determination of the inertial mass of an electron $m_e$ moving in a crystal lattice.\cite{Cabrera:89}

The negligible dissipation and approximate energy conservation of superconductors 
in the radio-frequency range give good conditions for the applicability of the Bernoulli theorem
\begin{equation}
\label{Bernoulli}
\frac12 m^*v^2n(T)+\rho_\mathrm{tot}\varphi=\rho_\mathrm{tot}\zeta,\quad
\rho_\mathrm{tot}=e^*n(t=0)
\end{equation}
for the constancy of the electro-chemical potential $\zeta$ in the condition of thermodynamic equilibrium with a constant temperature $T$,
cf.~the equations of hydrodynamics of a super-fluid by Landau.\cite{LL6}
If the superconductor is a clean metal with closed trajectories on the Fermi surface
$\rho_\mathrm{tot}=e(n_e-n_h)$ where $n_e$ and $n_h$ are densities of electronic states with positive and negative effective masses.\cite{LifshitzAzbelKaganov,LL10}

Let us recall that a voltmeter measures the difference of the electro-chemical potential.
That is why for the measurement of current induced contact potential difference
\begin{equation}
\label{Bernoulli_3D}
\Delta \varphi = -\frac12 m^*v^2n/\rho_\mathrm{tot}=-\beta j^2,
\qquad \beta(T)\equiv\frac{m^*/e^*}{2\rho_\mathrm{tot}^2\mathcal{C}_s(T)},
\qquad \mathcal{C}_s(T)\equiv\frac{\lambda^2(T=0)}{\lambda^2(T)}
=\frac{n(T)}{n(T=0)}\in (0,\,1),
\end{equation}
where $\Delta \varphi \equiv \varphi(v)-\varphi(0)$.
Introducing a dimensionless parameter $C_\lambda$ close to the critical temperature we have
\begin{equation}
C_\lambda \equiv -T_c\left.\frac{\mathrm{d}}{\mathrm{d}T}
\frac{\lambda^2(0)}{\lambda^2(T)}\right\vert_{T_c},
\qquad
\mathcal{C}_s(T)\approx C_\lambda\, |\epsilon|,
\qquad |\epsilon|=\frac{T_c-T}{T_c}\ll1,
\qquad T<T_c, 	
\qquad C_{\lambda,d}\approx 2.6,
\qquad C_{\lambda,s}\approx 2.0.
\end{equation}
For the experiment which we suggest it is indispensable to have capacitive coupling between the sample 
and the Bernoulli signal detector electrode. 
For a short review of the problem see Ref.~\onlinecite{Mishonov:94}
and references therein.
We wish to stress out that the Bernoulli theorem for superconductors is not a result of some dynamic theory but a consequence from thermodynamic consideration even of static super currents. 

Yet London\cite{London:35} pointed out, that superconducting alloys behave like
they have a negative surface tension\cite{Faber:52}
and Shubnikov\cite{Shubnikov:XX} even started to investigate the magnetization of the vortex phase
of now called type-II superconductors.\cite{Abrikosov_textbook}
Many years later in the framework of the Ginzburg-Landau theory\cite{GL,LL9}
an approximate dependence\cite{Mishonov:90} of the surface tension $\sigma$ from the 
Ginzburg-Landau (GL) parameter $\kappa$ was derived
\begin{equation}
\sigma(T)\approx \frac{8}{3\sqrt{2}}\,
p_B(T)\xi(T)\left(1-\sqrt{\frac{\varkappa}{\varkappa_c}}\right),
\quad p_B(T)=\frac{B_c^2(T)}{2\mu_0},
\quad \varkappa\equiv\frac{\lambda(T)}{\xi(T)}\approx \mathrm{const},
\quad \varkappa_c=\sqrt{2},
\quad \lambda(T)=\frac{\hat{\lambda}(0)}{\sqrt{|\epsilon|}},
\end{equation}
where $B_c(T)$ is the thermodynamic critical field related to the volume density
of the condensation energy $p_B(T),$
and $\hat{\lambda}(0)=\lambda(0)/\sqrt{C_\lambda(0)}$
parameterized penetration depth for $|\epsilon|\ll1$.
Here we wish to mention other well known results of GL theory 
\begin{align}
\varkappa= \frac{2^{3/2}\pi}{\Phi_0}\lambda^2(T)B_c(T),\qquad
\xi(T)=\frac{\xi(0)}{\sqrt{|\epsilon|}},\quad \epsilon=\frac{T-T_c}{T_c},
\quad
-T_c\left. \frac{\mathrm{d}B_{c2}}{\mathrm{d}T}\right\vert_{T_c}
=\frac{\Phi_0}{2\pi\xi^2(0)},
\qquad
\Delta C=\frac{T_c}{\mu_0}
\left( \left.\frac{\mathrm{d}B_{c}}{\mathrm{d}T}\right\vert_{T_c}\right)^2,
\nn
\end{align}
where $B_{c2}(T)\approx \Phi_0/2\pi\xi^2(T)$ is the super-cooling (for type-I superconductors) magnetic field,
$B_c$ is the thermodynamic magnetic field related to the jump 
of the heat capacity $\Delta C$ (per unit volume)
at the critical temperature $T_c$. 
The argument of $\xi(0)$ means only GL parametrization, but not $T=0$.

Unfortunately, theoretical considerations and brilliant physical intuition 
have never been taken seriously.
The charge of the superfluid charge carriers was not determined by the flux quantum
$e^*=2\pi\hbar [c]/\Phi_0$
an elementary consequence of the integer value 
of the dimensionless momentum circulation
$\oint (m^*\mathbf{v}+e^*\mathbf{A})
\cdot\mathrm{d}\mathbf{r}/2\pi\hbar.$
However scientific archaeology reveals that for first time idea
for electron doublets was proposed soon after discovery of superconductivity in 1914 by 
Sir~J.~J.~Thompson\cite{Thompson:14},
and later on the analyzing relation by superconductivity and Bose-Einstein condensation by Ogg\cite{Ogg:1946} and Shafroth.\cite{Schafroth:55}
In such a way it is clear why Onsager pointed out that $\Phi_0$ should 
correspond to $|e^*|=2|e|$ long time before the experimental determination.

All equations in the present work are written in almost system-invariant form;
as we pointed out in SI units $[c]$ must be substituted by one.
The research by Shubnikov on the now called Abrikosov vortex phase was even more dramatically interrupted.\cite{deGennes}
The same can be said for the effective mass of the super-fluid charge carrier $m^*$.
Analyzing only the temperature dependence of the penetration depth
$1/\lambda^2(T)=e^*\rho_\mathrm{tot}\mathcal{C}_s(T)/\varepsilon_0c^2m^*$ the mass of the Cooper pairs remains undetermined.
In all experiments with magnetic field only the ratio $n(T)/m^*$ participates
and with appropriate re-normalization of the density for $m^*$ one can take the mass of the Sun $m_{\odot}$, cf. Ref.~\onlinecite{Tinkham2}.
In such a way one significant part of the physics of superconductivity remains undeveloped despite of 10$^5$ works published on this topic. 
In order to measure $m^*$ it is necessary to study the electric field effect in superconductors using atomically clean superconductor surfaces, which is the main technological difficulty.
If we ignore Bernoulli and London-Hall effect in superconductors
we have one parametric re-normalization $n(T)\rightarrow C\,n(T)$ 
and simultaneously 
$m^*\rightarrow C\,m^*$, 
which conserves penetration depth ratio
\begin{equation}
\frac{n(T)}{m^*}=\frac{C\,n(T)}{C\,m^*}, \quad C>0
,\end{equation}
but this does not mean that $m^*$ is not accessible.
The density of super-fluid charge carriers can be determined also by measurement 
of the drift velocity of the  super-fluid by Doppler shift of plasmon resonances in thin films,
for example Ref.~\onlinecite{Mishonov:1994b}.
For clean superconductors half of the Cooper pair mass must be the extrapolated to zero frequency optical mass.\cite{Klenov:2003}

The purpose of the present  work is to suggest a simple experimental set-up and a method for investigation of the Bernoulli effect in superconductors, which is parameterized by the effective mass of Cooper pairs. 
In such a way it is possible to continue the first determination of the
Cooper pair mass\cite{Comment,Reply} begun in the Bell labs and
interrupted by the end of these laboratories.
In the next section we describe the suggested experiment in details.

\section{Description of the set-up and notations}%

Thin films with thickness $d_\mathrm{film}<\lambda$ grown on an insulator substrate are important objects for investigation in the physics of superconductivity.
If through such a film a two dimensional current $j_\mathrm{2D}$ 
with an acceptable 5\% accuracy flows one can consider that the bulk current density 
is homogeneous across the thickness of the film $j\approx j_\mathrm{2D}/d_\mathrm{film}$
and precisely this current density is on the interface of the superconductor with 
protecting insulator layer with thickness $d_\mathrm{ins}.$
We consider a superconducting film grown on a substrate and protected by a very thin insulator layer.
We suppose that the interface is perfect and the surface of the superconductor has the 
properties of the bulk material.

\begin{figure}%
\includegraphics[scale=0.5]{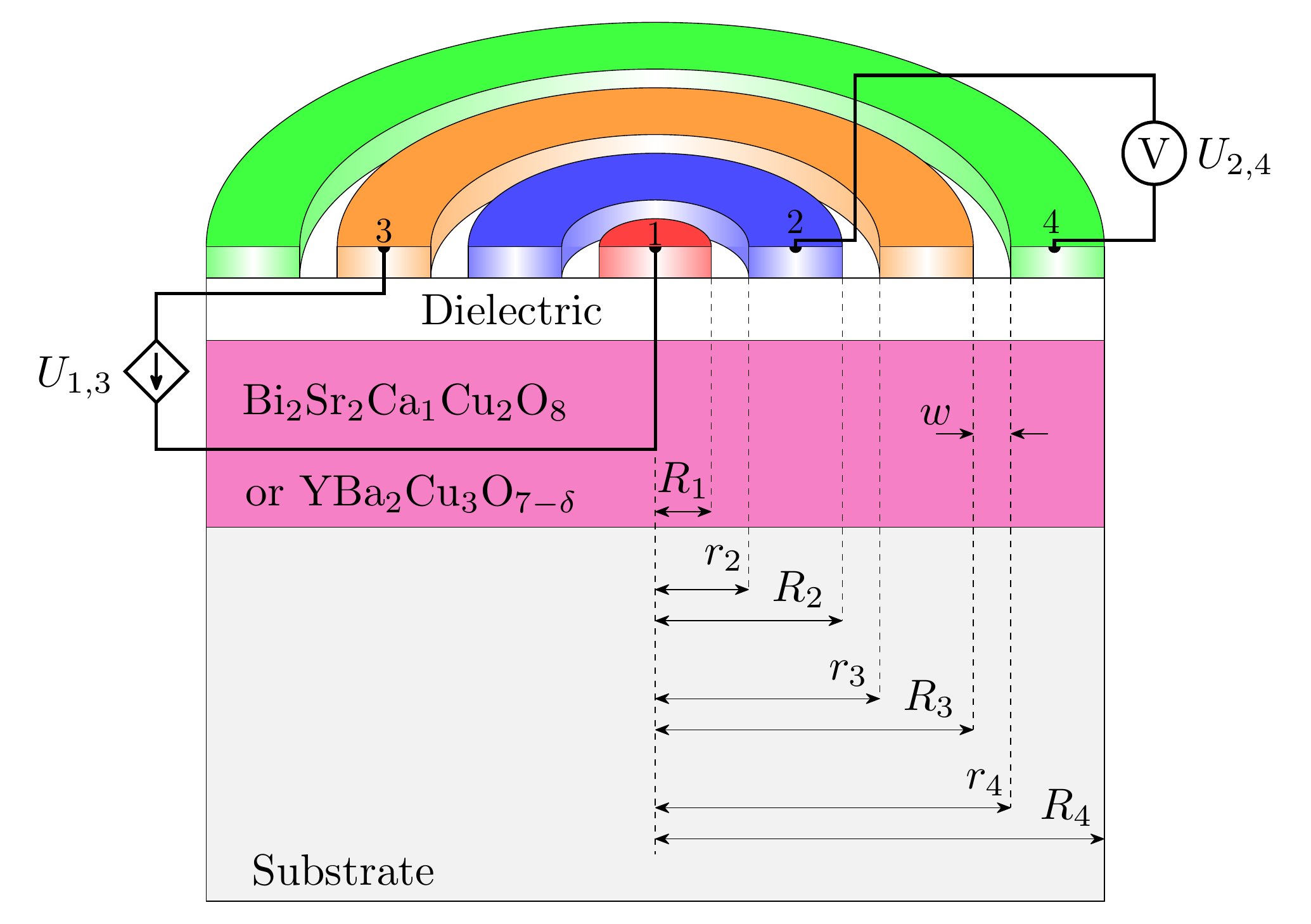}
\caption{Schematic presentation of hybrid structure for observation of the Bernoulli effect at the surface of a superconductor. 
The superconductor layer is grown on a square substrate.
An insulating layer is carefully prepared in order the superconductor 
surface to have almost bulk properties.
On the insulator layer 4 concentric metal ring electrodes are evaporated.
Drive voltage probes (1) and (3) are connected to a voltage generator.
Drive current passes in the superconductor layer below the electrode (2).
There is no current below the reference electrode (4). 
Current induced contact potential difference $U_{2,4}$
is measured by a capacitive coupling to the superconductor layer.
The drive and detector circuits are capacitively disconnected.
One dimensional version of this setup is depicted in Fig.~\ref{Fig:Wire}.
}
\label{Fig:Set-up}
\end{figure}

\begin{figure}%
\includegraphics[scale=0.25]{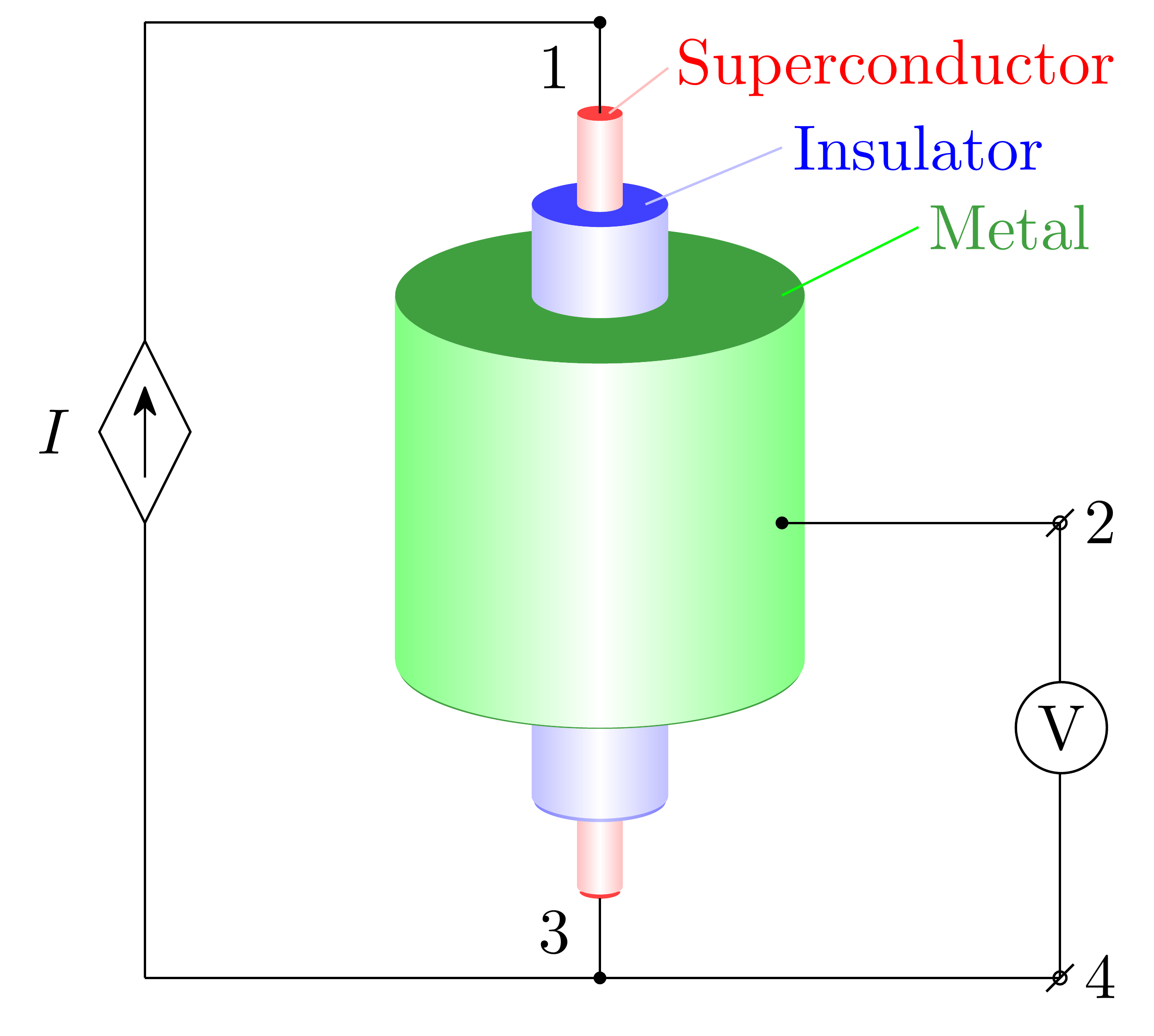}
\caption{One dimensional variation of the set-up from Fig.~\ref{Fig:Set-up}.
AC current generator in the left creates current $I(t)$ through the superconducting wire 
-- the narrow cylinder between points (1) and (3). 
The superconducting wire with radius $r$ is surrounded by an insulator (Insulator) and
there is a normal metal layer around the insulator (Metal).
As a whole we have a coaxial cable with a superconducting wire.
The detector circuit measures the AC Bernoulli voltage between 
probes (2) and (4). 
Comparison with Fig.~\ref{Fig:Set-up} shows that formally we have
infinite capacities  $C_1$, $C_3$ and $C_4$ and
there is a common point between the drive and detector circuits.
Practically we have a co-axial cable with a superconducting wire and normal metal around it separated by an insulator.
}
\label{Fig:Wire}
\end{figure}

\begin{figure}%
\includegraphics[scale=0.5]{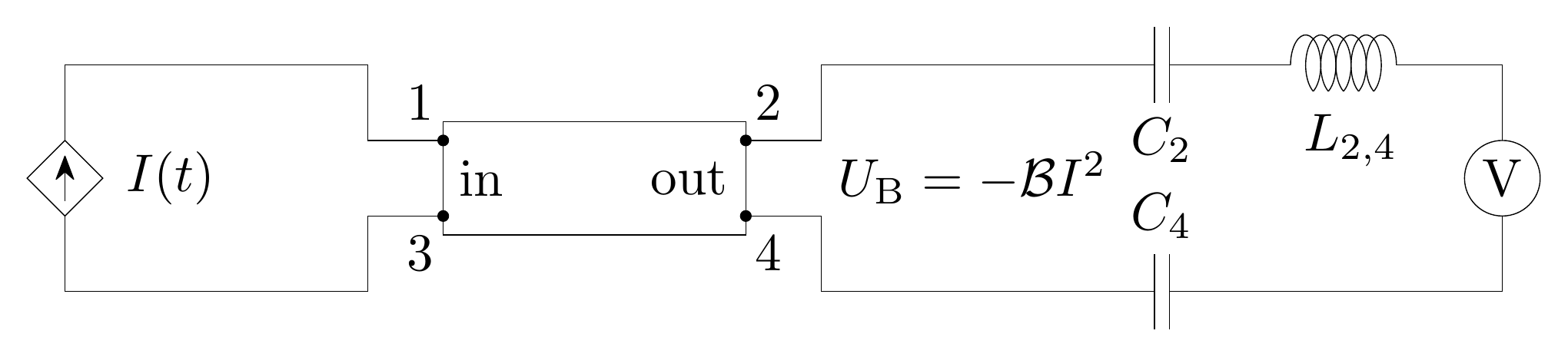}
\caption{An effective circuit for observation of the Bernoulli effect in superconductors.
A current generator creates a current $I(t)$, which is input for the two-port network.
The output voltage $U_\mathrm{B} = -\mathcal{B} I^2$ is detected through two capacitors with a total capacitance $C_2 C_4/(C_2+C_4)$.
The pre-amplifier and the lock-in voltmeter are shown schematically as a voltmeter.}
\label{Fig:Circ}
\end{figure}

On the protecting insulator layer 4 metal electrodes 
with axial symmetry  have to be evaporated, the material of the alloy is irrelevant.
In this Corbino geometry we have
1) One circle with radius $R_1$
2) Then after some narrow gap with width $w$ we have a ring electrode
with internal radius $r_2=R_1+w$ and external radius $R_2.$
This is repeated two times again with two other electrodes with radii
3) $r_3=R_2+w$ and external radius $R_3$, and finally
4) a ring electrode with internal radius $r_4=R_3+w$ and an external 
electrode with radius $R_4=a/2$ equal to the half side of the square substrate,
for definiteness let $a=5\,$mm.
All probes are wired and AC input drive voltage $U_{1,3}(t)$ is applied between
electrodes (1) and (3) and the time dependent output Bernoulli signal $U_{2,4}(t)$ is measured between electrodes (2) and (4).
This structure is depicted in Fig.~\ref{Fig:Set-up}; 
compare with one dimensional modification in Fig.~\ref{Fig:Wire}.
Both set-ups from Fig.~\ref{Fig:Set-up} and Fig.~\ref{Fig:Wire} have equivalent
electric circuit depicted in Fig.~\ref{Fig:Circ}.
%

The radii have such proportion that areas of the electrodes are approximately equal
\be
S_1=\pi R_1^2,\qquad
S_2=\pi (R_2^2-r_2^2), \qquad
S_3=\pi (R_3^2-r_3^2), \qquad
S_4=\pi (R_4^2-r_4^2).
\ee
The capacities neglecting the effect of ends are approximately 
$C_i=\varepsilon_0\varepsilon_\mathrm{ins}S_i/d_\mathrm{ins},$
where $i=1,2,3,4.$

The observable Bernoulli potential is determined by the super-fluid velocity and current density 
$j_\mathrm{surf}$ at the surface 
of the superconductor. 
This current should be much smaller than the critical one 
$j_c(T)$.
For thin superconductor film we have
\be
\Delta\varphi=-b_\mathrm{thin}j_\mathrm{2D}^2,
\quad b_\mathrm{thin}
=\frac{\beta}{d_\mathrm{film}^2}
=\frac{e^*}{2m^*}
\left(\frac{Z\lambda(0)\lambda(T)}
{\varepsilon_0c^2\,d_\mathrm{film}}\right)^{\!\! 2}
=\frac{m^*Z^2(x)}
{2e^*\rho_\mathrm{tot}^2d_\mathrm{film}^2\mathcal{C}_s(T)},
\quad 
Z(x)\equiv\frac{x}{\tanh x},
\quad x\equiv\frac{d_\mathrm{film}}{2\lambda}
.\ee 
Here the screening factor $Z(x\ll1)\approx1$
describes more precise treatment of the current density
$j\propto \exp(z/\lambda)$ across the thickness of the layer
$|z|<d_\mathrm{film}/2.$

Two dimensional current density 
$j_\mathrm{2D}=I/2\pi r$ is proportional to the total drive current $I$ between the electrodes (1) and (3).
The Bernoulli voltage is just the averaged Bernoulli 
potential beneath the reference electrode (2)
\be
U=\left<\Delta \varphi\right>_2=- \mathcal  B_\mathrm{thin}I^2,
\qquad \mathcal B_\mathrm{thin}
=\frac{b_\mathrm{thin}}{(2\pi)^2}
\Big\langle\frac1{r^2}\Big\rangle_{\!2},\qquad
\Big\langle \frac{1}{r^2}\Big\rangle
=\dfrac{\int_{r_2}^{R_2} \dfrac{1}{r^2} d(\pi r^2)}{\int_{r_2}^{R_2}  d(\pi r^2)}
=\dfrac{2\ln\dfrac{R_2}{r_2}}{R_2^2-r_2^2},
\ee
or finally
\be
\label{Bernoulli_thin}
\mathcal B_\mathrm{thin}
=\frac{e^*}{4\pi^2m^*}
\frac{\ln\frac{R_2}{r_2}}{R_2^2-r_2^2}
\left(\frac{Z\lambda(0)\lambda(T)}
{\varepsilon_0c^2\,d_\mathrm{film}}\right)^{\! 2}
=\frac{m^*}{e^*}
\frac{Z^2\ln(R_2/r_2)}
{(2\pi\rho_\mathrm{tot}d_\mathrm{film})^2\,
(R_2^2-r_2^2)\,\mathcal{C}_s(T)}
.\ee 
For thin film $d_\mathrm{film} \ll 2 \lambda (T)$, $Z \approx 1$.
In the first expression it is supposed that we know
the experimentally determined temperature dependent penetration depth,
while in the second one is supposed 
that we know the total volume density of the charge carriers
and the temperature dependent relative super-fluid concentration $\mathcal{C}_s(T).$
Those expressions are simplified at temperatures much lower than
the critical one $T\ll T_c$ and very thin layers 
$d_\mathrm{film}\ll \lambda(0)$
when we have $Z=1=\mathcal{C}_s(0).$
As $|e^*|=2|e|$ the measurement of the Bernoulli voltage 
$U_{2,4}=U$, see  Fig.~\ref{Fig:Set-up}, determines the Bernoulli coefficient 
$B_\mathrm{thin}$ and the effective Cooper pair mass $m^*$.

In order to clarify the work of the set-up in the next section we will analyze the case of a bulk superconductor.

\section{Bernoulli effect on the surface of a bulk superconductor}
Now let us describe the work of the set-up when the insulator layer and electrodes are deposited on the
atomically clean surface of a bulk superconductor.
Imagine a good Bi$_2$Sr$_2$Ca$_1$Cu$_2$O$_8$ crystal 
cleaved by adhesive tape.
In this case the volume density of the current decreases
exponentially in the bulk of the superconductor 
$j(z)\propto \exp(-z/\lambda),$
and the current density on the surface 
$j_\mathrm{surf}=j_\mathrm{2D}/\lambda$ expressed by the two dimensional current density and the temperature dependent penetration depth.
At every two dimensional space point $\boldsymbol\rho=(x,y)$
surface current density is proportional to the drive current
$I\equiv I_{1,3}$ between electrodes (1) and (3)
\be
j(\boldsymbol\rho)
=I/S_\mathrm{bulk}(\boldsymbol\rho),
\qquad S_\mathrm{bulk}(\boldsymbol\rho)=2\pi r \lambda.
\ee
For thin $d_\mathrm{film}\ll\lambda$ films  
the geometric factor
$S_\mathrm{thin}=2\pi r d_\mathrm{film}$
is the area of a thin cylindrical slice.
For the bulk sample we have
\be
\label{bulk}
\Delta \varphi=-b_{\mathrm{bulk}}\,j_\mathrm{2D}^2
=\frac{B^2}{2\mu_0\rho_{\mathrm{tot}}},
\qquad
b_{\mathrm{bulk}}\equiv \frac{\beta}{\lambda^2} = \frac{1}{2\epsilon_0 c^2 \rho_{\mathrm{tot}}},
\qquad 
B=\mu_0 j_\mathrm{2D}\le B_c(T),
\ee
where $B$ is parallel to the surface magnetic field 
created by the two dimensional current.
Averaging again the $1/r$ dependence of the current density under the Bernoulli electrode (2) we obtain in the same manner the Bernoulli constant for 
the set-up with a bulk sample
\be
U=- \mathcal  B_\mathrm{bulk}I^2,\qquad
\mathcal  B_\mathrm{bulk}=
\Big\langle \frac{b(\mathbf{\boldsymbol\rho})}{{S^2(\mathbf{\boldsymbol\rho})}} \Big\rangle_{\! 2}
=\frac{\ln(R_2/r_2)}{4\pi^2(R_2^2-r_2^2)\,\varepsilon_0c^2\rho_\mathrm{tot}}.
\ee
Let us mention the lack of the temperature dependence in the Bernoulli potential for the bulk sample.
One can consider that Bernoulli effect in this case is London
Hall effect describing how the magnetic pressure
$p_B=B^2/2\mu_0=\rho_\mathrm{latt}\,\Delta\varphi$
is transmitted to the ion lattice with volume density of charge $\rho_\mathrm{latt}=-\rho_\mathrm{tot}.$
Recalling $\Delta \phi =- \int E_z(z) \,\md z$
we can write the volume density of the force acting on the lattice as
\begin{equation}
\mathbf{f}_\mathrm{latt}
= \rho_\mathrm{latt} \mathbf{E}=\mathbf{j}\times\mathbf{B}
=-\nabla p_{\!B}
=  \nabla\cdot\frac1{\mu_0} \left(
\mathbf{B}\mathbf{B}-\frac12 B^2\openone
\right),
\end{equation}
i.e. the Lorentz force is gradient of the magnetic pressure and 
divergence of the Maxwell stress tensor.~\cite{LL2,LL7}
Those 3 vectors are mutually orthogonal
$E_z=R_\mathrm{LH} \,B_\varphi \, j_r$,
where $R_\mathrm{LH}=1/ \rho_\mathrm{tot}$
is the not yet ever measured constant of the London Hall effect
written in SI units. 
Let us remark the original idea by London:
for the surface of a bulk crystal 
a change of the potential 
depends on the tangential to the surface magnetic field
$\Delta \varphi = R_\mathrm{LH} \,B_\mathrm{t}^2/2\mu_0.$
We would like to add that this Bernoulli potential or also the London Hall effect potential is temperature independent, while in Ref.~\onlinecite{Greiter:89} the temperature dependent density of the superfluid is used in their final Eq.~(3.23).
The reason for this difference is that in Eq.~(\ref{Bernoulli}) the kinetic energy is proportional to the superfluid density, while the potential energy includes the total charge density of both the superfluid and normal charge carriers.

In such a way measurement of the Bernoulli
voltage $U_{2,4}=U$ leads to direct determination
of the total density of charge carriers $\rho_\mathrm{tot}$.
This important parameter can be substituted in the formula 
for the Bernoulli coefficient for a thin film 
Eq.~(\ref{Bernoulli_thin})
to determine the effective mass of Cooper pairs $m^*$.
The comparison gives
\be
\frac{\mathcal B_\mathrm{thin}(T)}{\mathcal B_\mathrm{thick}}
=\left(\frac{Z(T)\lambda(T)}{d_\mathrm{film}}
\right)^{\! 2},\ee
and we have obtained a new method for determination 
of penetration depth $\lambda(T)$ in type-II superconductors.
In order to obtain the derived formulas in Gaussian units
we have to replace $\Phi_0\rightarrow \Phi_0/c$ and to substitute $\varepsilon_0=1/4\pi$ and 
$\mu_0=4\pi.$
In Heaviside-Lorenz system we have to substitute
$\varepsilon_0=1$, $\mu_0=1$ and $c=1$.

Performing analogous calculations for a superconducting 
wire with radius $r$,
i.e. for the ``coaxial cable'' set-up shown in 
Fig.~\ref{Fig:Wire} we have for thin $r\ll \lambda$
and thick wire $r\gg \lambda$
\be
\mathcal B_\mathrm{thin}^\mathrm{(wire)}
=m^*/2e^*\rho_\mathrm{tot}^2\mathcal C_s(T) (\pi r^2)^2 =
R_\mathrm{LH} \lambda^2(T)/2 \varepsilon_0 c^2 (\pi r^2)^2,
\qquad
B_\mathrm{thin}^\mathrm{(wire)}
/B_\mathrm{thick}^\mathrm{(wire)}
=(2\lambda(T)/r)^2.
\ee
The general expression containing Bessel functions 
can be easily programmed for the experimental data processing.

\section{Experimental method}%
Let apply to electrodes (1) and (3) driving voltage, which is a sum of two sinusoidal,
one basic and one modulated with much smaller frequency 
\begin{equation}
U_{1,3}(t)
=U_a\sin(\Omega t)+U_b\sin(\Omega t)\sin(\omega t)
=U_a\sin(\Omega t)
+\frac12 U_b\,[\cos((\Omega-\omega)t)-\cos((\Omega+\omega)t)],
\qquad \omega\ll\Omega.
\end{equation}
Then through these electrodes a current
\begin{align}&
I_{1,3}(t)=C_{1,3}\mathrm{d}_t U_{1,3}
=I_a\cos(\Omega t)-\frac12\,I_b\sin((\Omega-\omega)t)
+\frac12\,I_b\sin((\Omega+\omega)t),\\
&
I_a\approx\Omega\, C_{1,3} \,U_a,
\qquad 
I_b\approx\Omega\, C_{1,3}\, U_b,
\qquad C_{1,3}\equiv\frac{C_1C_3}{C_1+C_3}
\end{align}
flows.
This current creates Bernoulli voltage
\begin{align}
&U_{2,4}(t)=-\mathcal BI_{1,3}^2
=-\frac12 \mathcal B\, I_a I_b \sin(\omega t)+\dots,
\nonumber
\end{align}
where the high frequency terms with frequencies $2\Omega\pm\omega$
are not written.
The usage a low noise pre-amplifier should be followed by a selective resonance amplifier tuned to $f=\omega/2\pi$.
In this case if we have monochromatic Bernoulli voltage
\begin{equation}
U_{2,4}(t) \approx - U_\mathrm{B}\sin(\omega t),
\qquad
U_\mathrm{B}\equiv\frac12 \mathcal B I_aI_b.
\end{equation}
This AC voltage with amplitude $U_\mathrm{B}$ has to be measured 
as connected in series to a capacitor with capacitance
$C_{2,4}\equiv C_2C_4/(C_2+C_4)$.
In short, we need a high-frequency signal with frequency $\Omega$,
which is modulated with low frequency $\omega$, and the same 
low-frequency signal is applied to the lock-in as reference signal.

If the thickness of the insulator $d_\mathrm{ins}$ is not extremely thin,
the corresponding impedance $1/\omega C_{2,4}$ is very high
and this creates significant difficulties for the measurements.
The Bernoulli signal must be bi-linear with respect to $U_a$ and $U_b$.
However the detector circuit can have its own non-linearity. 
This background can be measured above $T_c$ and is reliably 
removed by the temperature dependence of the Bernoulli signal.
The suggested experiment is actually not difficult.
The experimental technique gave the possibility for 
observation of the Bernoulli signal long time ago; 
confer the pioneer works by Lewis,\cite{Lewis}
Bok and Klein,\cite{BokKlein:68} and Morris and Brown,\cite{MorrisBrown:71}
simply the importance 
of the Bernoulli effect in superconductors was not evaluated at those times.
Material science of the superconductors was not on the agenda of the condensed matter physics and the measurement of the Cooper pair mass $m^*$ was not related to the Bernoulli effect.

As we demonstrate precise measurements of the Bernoulli
constant for thin film $\mathcal B_\mathrm{thin}(T)$ 
and bulk sample $\mathcal B_\mathrm{bulk}$ give simultaneously
the effective mass of Cooper pairs $m^*$,
the total volume charge density of the charge carriers 
$\rho_\mathrm{tot}$, and the temperature dependent penetration depth $\lambda(T)$
\be
\rho_\mathrm{tot}
=\Big\langle \frac{b(\mathbf{\boldsymbol\rho})}{{S^2(\mathbf{\boldsymbol\rho})}} \Big\rangle_{\! 2}
=\frac{\ln(R_2/r_2)}{4\pi^2(R_2^2-r_2^2)\,\varepsilon_0c^2B_\mathrm{bulk}},
\qquad
\lambda(T)
=\frac{d_\mathrm{film}}{Z(T)}
\sqrt{\frac{\mathcal B_\mathrm{thin}(T)}
{\mathcal B_\mathrm{thick}}},
\qquad
m^*=\frac{e^*\rho_\mathrm{tot}\lambda^2(0)}
{\varepsilon_0c^2}.
\ee
Here we have to use some extrapolated to zero temperature value using the smooth temperature function 
$\mathcal{C}_s(T)=\lambda^2(0)/\lambda^2(T)$.

The Bernoulli equation Eq.~(\ref{Bernoulli}) is applicable for currents much smaller than the maximal current density\cite{Tinkham2} in Ginsburg-Landau approximation for $|\epsilon|\ll1$
\begin{equation}
j_\mathrm{max}(T)
\approx\frac{2}{3\sqrt{3}}\frac{\varepsilon_0c^2\,\hbar/|e^*|}{\xi(T)\lambda^2(T)}=
\frac{2 \sqrt{2}}{3 \sqrt{3}} \frac{B_c(T)}{\mu_0 \lambda(T)},
\qquad
B_c(T)\approx\frac{1}{2\sqrt{2}\,\pi}\frac{\Phi_0}{\xi(T)\lambda(T)},
\qquad
B_{c1}(T)\approx\frac1{2\pi}\frac{\Phi_0}{\lambda^2(T)}\ln\varkappa,
\end{equation}
where the numerator in the first expression has dimension velocity times electric charge;
in Gaussian system we have to substitute $\mu_0=4\pi$
and $B_c(T)=cH_c(T);$
in the last expression $\varkappa\gg 1.$
For $j\ll j_\mathrm{max}$ the drift momentum of the  Cooper pairs
$m^*v\ll\hbar/\xi(T)$.
This GL result is derived minimizing the volume density of the Gibbs free energy
$g=(a(T)+p^2/2m*)n+bn^2/2$ with respect to the density $n$,
where for $T<T_c$ the coefficient $a(T)=-\hbar^2/2m^*\xi^2(T)=a_0\epsilon$
is negative. 
The impurity and disorder parameter of the GL coefficients $a_0$ and $b$
was calculated in Ref.~\onlinecite{Pokrovsky:2003},
the impurity dependence of the Cooper pair mass can be understood even in the framework of Pippard-Landau theory cf. Ref.~\onlinecite{Mishonov:1994b} 
If the current density is significant within the GL theory one can easily obtain
\begin{align}
& Q=m^*v \, \xi(T)/\hbar \leq 1,
\qquad p(t)=m^*v(t)=e^*\int_0^t E(T) \, \mathrm{d}t,
\qquad j=e^* v \, n(T,v),\qquad v(t=0)=0,\\
&\frac{j}{j_\mathrm{max}}=f(Q),\qquad f(Q)=\frac{3\sqrt{3}}{2}Q(1-Q^2)\leq 1, \qquad
-\frac{\Delta \varphi}{\beta j^2}=g(Q), \qquad g(Q)=\frac{1}{1-Q^2}.
\end{align}

Needless to say that in the usual BCS theory the superconducting gap $\Delta(T)$
is just the order parameter which minimizes the density the free energy $g$ in the self-consistent BCS calculations. 
BCS gap is accessible by tunneling spectroscopy and far-infrared spectroscopy as well.
In static GL theory, with static effective wave function $\Psi(\mathbf{r})$
and static vector-potential $\mathbf{A}(\mathbf{r})$ only the ratio $|\Psi|^2/m*$ can be determined.
In order to determine the Cooper pair mass it is necessary to study electrostatic
Bernoulli potential or London Hall effect in bulk superconductors.
Indeed London analyzed this physical situation 70 years ago and the
Bernoulli potential in superconductors has been analyzed in the framework
of BCS theory many years ago.\cite{Hong}
In general, the methods of the statistical physics allow us to analyze every situation 
with many particles.
For conventional clean superconductors $m^*$ is just extrapolated to 
zero frequency optical mass.

The goal of the present work is to point out that Cooper pair mass $m^*$
can be measured in every laboratory involved in the research of superconductivity.
In the next section we will give a detailed numerical example which will
illustrate the order of the value of the predicted effect.

Here we present only one opportunity to explore electric field effects for
revealing fundamental properties of superconductors.
We consider that electrostatic excitation of the currents is the best method 
for thin films and two dimensional superconductors.
Another possibility for creation of Cooper pair mass spectroscopy 
by current induced contact potential difference is to use
magnetic field for creation of eddy currents as it is for the standard method for
mutual inductance measurements.\cite{Fiory:88}
We believe that this method is better for thick films and bulk crystals.
Another obvious possibility is to use permanent magnet
driven by piezoelectric crystal oscillations.

\section{Numerical evaluation of the Bernoulli effect}%
Quadratic variations of the chemical potential as function of current density was considered by
London and Landau but  the experiment has not yet been done.
In order to urge experimentalists to explore fundamental properties of superconductors using Bernoulli effect perhaps it is necessary to give an illustrative numerical example in 
order to evaluate that the experiment is doable.
Let us take numerical values similar to Bi$_2$Sr$_2$Ca$_1$Cu$_2$O$_8.$
Let us accept $m^*=10 m_e$ and estimate the voltage which has to be measured.
Let take $T_c=88\,\mathrm{K}$ and working temperature which is $\Delta T=$3~K below the critical one, $T=85\,\mathrm{K}$. 
For the thickness of the superconductor layer we choose 
$d_\mathrm{film}=100\,\mathrm{nm}.$
The insulator layer should be actually very thin 
with effective thickness divided by dielectric constant
$d_\mathrm{ins}=10\,\mathrm{nm}.$
The normal metal layer in this hybrid structure is irrelevant but for very thin Al layers we can use
self-healing of the pinholes due to evaporation of the metal around.
Let us accept also material constants: 
GL extrapolated penetration depth
$\lambda(0)=260\,\mathrm{nm}$ 
and coherence length
$\xi(0)=26\,\mathrm{nm}$; $\varkappa=62.0$.
For this superconductor with d-symmetry of the superconducting gap
$C_{\lambda,d}=2.6$ which gives
GL parameterizing of the penetration depth
$\hat{\lambda}(0)=161\,\mathrm{nm}$ and 
$\varkappa= 62$.
For these conditions we have: $|\epsilon|=0.0341$, 
$\lambda(T)=873\,\mathrm{nm}\gg d_\mathrm{film}$,
$Z\approx 1$,
$\xi(T)=140.8\,\mathrm{nm}$,
$\mathcal{C}_s= 0.0886$.
We calculate also 
$B_{c1}(T)=994\,\mu\mathrm{T},$
$B_c(T)=59.4\,\mathrm{mT},$
$B_{c2}(T)=1.660\,\mathrm{T},$
$j_\mathrm{max}(T)=29.49\,\mathrm{GA/m^2},$
$\beta=1.432\times 10^{-27}\,\mathrm{V A^{-2} m^4},$
$\mathcal{B}_\mathrm{thin}(T)=399\,\mathrm{pVA^{-2}}.$

Let have a $10\times10\,\mathrm{mm}$ sample of
substratum, epitaxial Bi$_2$Sr$_2$Ca$_1$Cu$_2$O$_8$,
and thin dielectric layer.
On the insulator layer 4 concentric metalic probes  are evaporated:
a central circle with radius $R_1=2.50\,\mathrm{mm}$,
and 3 ring electrodes with external radii 
$R_2=3.54\,\mathrm{mm}$,
$R_3=4.33\,\mathrm{mm}$,
and $R_4=5.00\,\mathrm{mm}.$
The gap between electrodes has width $w=20\,\mu\mathrm{m},$
and the internal radii are $r_2=R_1+w$, $r_3=R_2+w$ and $r_4=R_3+w$;
see Fig.~\ref{Fig:Set-up}.
So that areas of all plane capacitors are approximately equal and
$C_{1,3}=8.59$~nF and $C_{2,4}=8.50$~nF.

The basic sinusoidal voltage with frequency $\Omega/2\pi=250$~MHz 
has amplitude $U_a=500$~mV
and the modulated with low frequency $\omega/2\pi =250$~kHz signal has
the same amplitude  $U_b=500$~mV.
For the brake-down electric field of the insulator we take
$E_\mathrm{bd}=1$~GV/m.
The maximal electric field in the insulator layer we estimate as
$E_\mathrm{max}=(U_a+U_b)/d_\mathrm{ins}= 100\,\mathrm{MV/m}$,
so that $E_\mathrm{max}/E_\mathrm{bd}=0.1$ and 
maximal charge on the sequential capacitors $C_1$ and $C_3$
is $Q_{1,3}=8.59\,\mathrm{nC}$.

At high-frequencies $\Omega$ the capacitive impedance between probes (1) and (3)
is $Y_{1,3}=1/(\Omega \,C_{1,3})=74\,\,\mathrm{m\Omega}$
and the corresponding currents are significant
$I_a=U_a/Y_{1,3}=6.75$~A and 
$I_b=U_b/Y_{1,3}=6.75$~A is the same.
The maximal two dimensional current density is
$j_\mathrm{2D}=(I_a+I_b)/(2\pi r_2)=852$~A/m,
and the bulk one
$j=j_\mathrm{2D}/d_\mathrm{film}=8.52\,\,\mathrm{GA/m^2}$;
$j/j_\mathrm{max}(T)=0.289.$
The super-fluid drift velocity we evaluate at $v=j/e^*n_\mathrm{tot}\mathcal{C}_s(T)=204$~m/s,
which is smaller that de-pairing current at this temperature
$v_\mathrm{dep}(T)=\hbar/m^*\xi(T)=822$~m/s.
The dimensionless momentum is $q=m^*v\,\xi(T)/\hbar=0.249$.
Let us mention also the extrapolation 
$v_\mathrm{dep}(0)=\hbar/m^*\xi(0)=4.45$~km/s.

One can evaluate now the amplitude of the Bernoulli voltage 
$U_\mathrm{B}=9.09$~nV,
which after amplification $A=10^6$ becomes
$U_\mathrm{ampl}=A U_\mathrm{B}=9.09$~mV.
For the calibration of $A=10^6$ amplifier is convenient to use
shot\cite{epo6} or thermal noise.\cite{epo5}

The rectified signal is collected on the lock-in capacitors for, say, $\Delta t=5$~s.
Evaluating the electric voltage noise of the first operational amplifiers of the 
pre-amplifier to have spectral density parameter 
$e_\mathrm{N}=1.2,\mathrm{nV/\sqrt{Hz}}$,
typical noise voltage can be evaluated as 
$U_\mathrm{N}=e_\mathrm{N}\sqrt{\Delta t}=2.68$~nV
we obtain Bernoulli signal to noise ratio
$r_\mathrm{S/N}=3.39$,
which reveals that experiment is doable,
and Cooper pair mass in cuprates can be determined using standard electronic equipment.

In the considered conditions the Bernoulli effect operates as a de-modulator and
one of the difficulties is that the small nano-Volt signal has to be measured
hidden in the capacitive coupling with impedance 
$Y_{2,4}=1/(\omega C_{2.4})= 74.9\,\,\Omega.$
This capacitive impedance can be compensated with an inductance
$L=1/\omega^2 C_{2,4}= 47.6\,\,\mathrm{\mu H}.$
Using general impedance converter this inductance can be tunable.
A Python script for calculation of this numerical illustration is given in the appendix.

This example demonstrates also the possibilities for different modification.
Indispensable is only $C_2$ capacitor.
All other electrodes can be soft gold electrodes directly galvanically touched to the surface of the superconductor. 
In this case breakdown problem of the insulator layer can be avoided.
The central gold circle probe (1) and gold ring probe (3) can be terminals of a coaxial cable.
Similar system was used for measurement of penetration depth in thin films\cite{Anlage};
here we suggest a three-axial version of this device, which can give also the Cooper pair mass.

For Bernoulli effect we have to apply strong current and to measure small electric voltage.
One electric field effect in superconductors is in some sense complementary.
One can apply strong electric field and to measure small magnetization related to eddy currents.
We analyze it in the next section.

\section{Electric field modulation of the kinetic inductance, 
condensation energy and the condensation energy}%

The order of the Bernoulli effect in high-$T_c$ cuprates leads to very optimistic
evaluation of the difficulties of the sample preparation. 
However the causes of the  several unsuccessful attempts with YBa$_2$Cu$_3$O$_{7-\delta}$ perhaps was the degradation of the surface 
(CuO$_2$)$_2$ bi-layer by vapor. 
It is necessary to use carefully prepared interfaces for investigation of other field effects in superconductors. 
The considered structure can be studied by two coil mutual inductance method,
where the mutual inductance  is parameterized by the kinetic inductance
\begin{equation}
L(T) = \frac{m^{\star}}{{e^{\star 2}} n_\mathrm{2D}(T)} = \mu_0 \frac{\lambda^2(T)}{d_{\mathrm{film}}},
\qquad n_\mathrm{2D}(T)=d_\mathrm{film}n(T),
\end{equation}
which is determined by two dimensional density 
$n_\mathrm{2D}(T)$ of the superfluid particles. 
The Bernoulli coefficient $b_\mathrm{thin}$ can be expressed 
by the kinetic inductance $L(T)$
\begin{equation}
\Delta\varphi=-b_\mathrm{thin}j_\mathrm{2D}^2,\qquad
b_{\mathrm{thin}}(T)
= \frac{L(T)}{2\rho_\mathrm{tot} d_{\mathrm{film}}},\qquad
\rho_\mathrm{tot}=\frac{L(T)}
{2 d_{\mathrm{film}}b_{\mathrm{thin}}(T)},
\end{equation}
and the ratio gives the temperature independent 
total charge density of charge carriers $\rho_\mathrm{tot}$.
For clean crystals this volume density $\rho_\mathrm{tot}$ describes 
the Hall effect in strong magnetic field.\cite{LL10}
\begin{figure}[h]%
\includegraphics[scale=0.75]{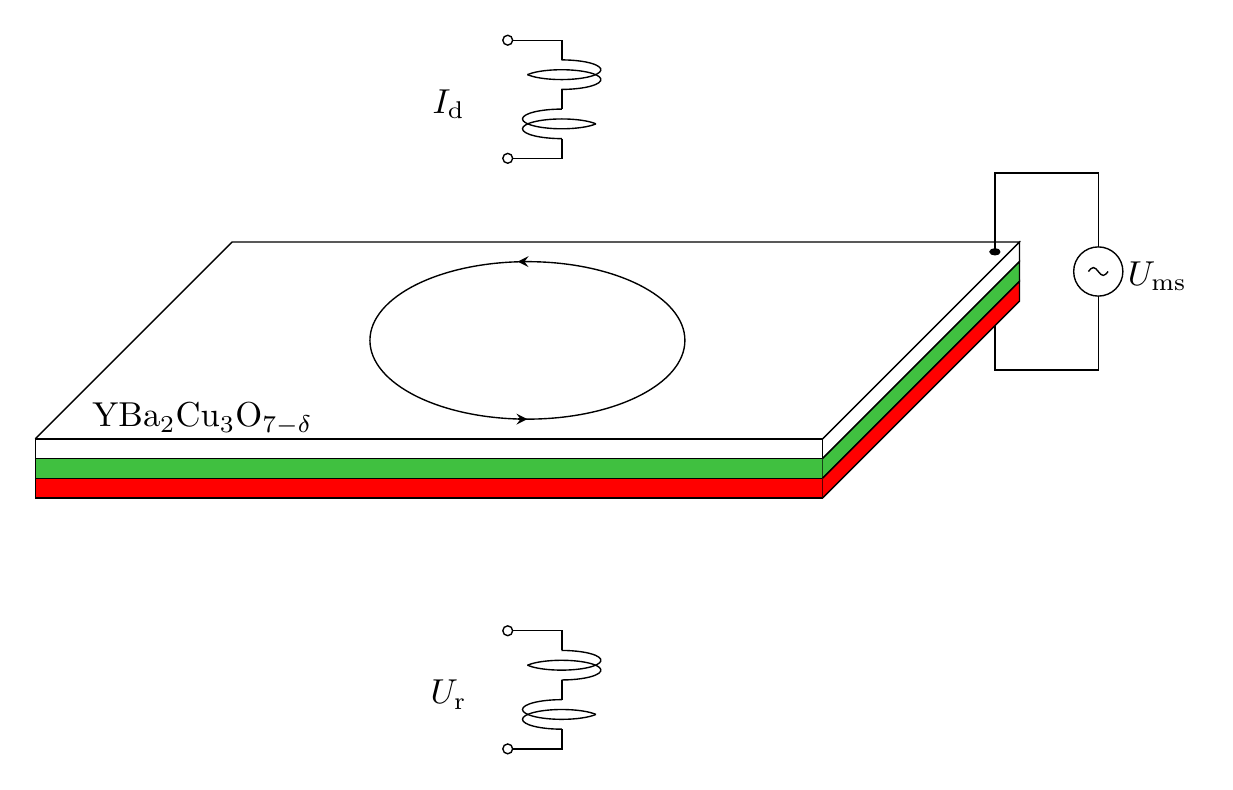}
\caption{Set-up for Cooper-pair mass spectroscopy by
electrostatic modulation of the kinetic inductance; 
after Ref.~\onlinecite{Fiory:88}. 
Current through drive coil $I_\mathrm{d}$ induces eddy current in the superconductor film which is detected as voltage $U_\mathrm{r}$ in detector circuit.
Modulating voltage $U_\mathrm{m}$ is applied between the superconductor layer and the metal electrode. 
Additional magnetic moment is created by excess superconducting charge carriers 
induced by electric induction in this plane capacitor on the superconductor-insulator interface. 
Several principally different experiments van be performed by
small modifications of this set-up.
}
\label{Fig:Fiory}
\end{figure}%

However the considered structure can be used for study of completely different effect:
electric field modulation of the kinetic inductance. 
At this effect a strong electric field is applied between superconducting an insulator layer 
which forms a plane capacitor with the surface charge density on the plates $Q=\varepsilon_0\varepsilon_\mathrm{ins}U_\mathrm{ms}/d_\mathrm{ins},$ 
where $U_\mathrm{ms}$ is the metal-superconductor voltage and 
$\varepsilon_\mathrm{ins}$ is the relative dielectric constant of the insulator layer.
The effective mass of Cooper pairs can be expressed\cite{Comment} 
by the ratio of the variations
\be
m^*=-2|e|\mathrm{sgn}(e^*) L(0)L(T)\frac{\delta Q}{\delta L(T)},
\qquad
\mathcal{C}_s(T)=\frac{L(0)}{L(T)}.
\ee
Actually this method has been used for first determination 
of $m^*\approx 11\, m_e$
in YBa$_2$Cu$_3$O$_{7-\delta}$ (Y:123).\cite{Reply}
This experimental set-up is depicted in Fig.~\ref{Fig:Fiory}.

As the quality of the interface is crucial for the suggested experiments 
the breakthrough perhaps will be triggered by the easily cleaved surface of 
Bi$_2$Sr$_2$Ca$_1$Cu$_2$O$_8$ (Bi:2212).
In short, Bi:2212 crystals good for 
ARPES (Angular Resolved Photo-Emission Spectroscopy)
and Bi:2212 films good for making squids are perfect for the 
pioneering observation of Bernoulli effect in superconductors.

The Eq.~(\ref{bulk}) describes the transmission of the magnetic pressure 
to the lattice. 
This equation can be used up to the critical thermodynamic field $B_c(T)$
and this gives a new method for determination of the thermodynamic critical field 
and condensation energy
\be \rho_{\mathrm{tot}} \left [\varphi(T)-\varphi(T_c)\right]=B_c^2(T)/2\mu_0\ee
related to the jump of the heat capacity.
The method requires the same capacitive junction to the sample.
The superconductor is illuminated by a chopped light in order the surface to 
become a normal metal. 
The amplitude of modulation of the contact potential difference 
is determined by the amplitude the induced charge detected by the electronic circuit.
In such a way the density of condensation energy $B_c^2(T)/2\mu_0$
can be measured by the electric measurements without the use of magnetic field.

\section{Discussion and conclusion}%

Perhaps the use of Josephson effect 
and the formula for the frequency created by a constant voltage 
$\omega_\mathrm{J}=e^*U/\hbar$ is the most famous 
electric field effect in superconductors.
The electric field could be static but we observe a dynamic effect,
roughly speaking because the electric potential has to participate in the gauge 
invariant equations only by $(\mathrm{i}\hbar\partial_t-e^*\varphi)\Psi$.
The effective mass of Cooper pairs $m^*$ is a typical parameter of low frequency dynamics 
of superconductors, even in the case of static electric field. 
But in any case the determination
of $m^*$ and $R_\mathrm{LH}$ requires electric fields 
(applied strong electric field or measured small electric voltage).
In the framework of purely static Ginzburg-Landau theory
with static order parameter $\psi(\mathbf{r})$ and only space dependent
vector-potential $\mathbf{A}(\mathbf{r})$ the effective mass of Cooper pairs 
$m^*$ cannot be determined.
This statement is trivial but created significant retention of the development of the physics of superconductivity.
Authors of grant proposals and referee reports were afraid that something new will be investigated.
From social aspect it is well-known that transparent theoretical ideas 
slowly receive experimental observation.
Flux quantization was predicted by London\cite{London:50} in 1950 but was 
wad observed 11 years after when $|e^*|=2|e|$ was trivialized by the BCS theory.
Absolutely analogously already 70 years $m^*$ and $R_\mathrm{LH}$ 
are still in the agenda of the physics of superconductivity,
and without the mass of Cooper pairs $m^*$ remains Hamlet without the Prince but only
with an onnagata in the role of Ofelia.

Electric field effects in superconductors were studied mainly in attempts to get electronic 
applications, however a lot of fundamental physics can be obtained if we carefully 
investigate cleaved superconductors or nano-structures with good quality of the interface layer.
It is necessary to have no degradation of the superconductivity in the last nano-meter layer under the interface.
We use understandable notions as the effective mass $m^*$ and volume charge density
$\rho_\mathrm{tot}$ just to parameterize the effects. 
However, the contact potential difference and the change of the superconducting properties by electrostatic charge modulation can be studied experimentally and calculated theoretically directly from the microscopic theory even without the help of the so called phenomenological parameters. 
Roughly speaking $m^*$ can be determined by thin films,
while for bulk samples one can measure $\rho_\mathrm{tot}$ 
and one can determine penetration depth $\lambda$ observing only Bernoulli effect in superconductors.

A lot of new physics can be obtained by investigating those well-forgotten effects.
The technical applications will come without a significant delay.
Let us only note that  the Bernoulli effect is quadratic for the frequencies up to the superconducting gap.
This quadratic effect can be used for creation of high frequency de-modulators 
and even TeraHertz lock-in voltmeters.
 

\section*{Appendix. The Python script for the numerical example and its results}

\begin{lstlisting}[language=Python,breaklines=true]
#!/usr/bin/python
print; print; print;print; print; print
print "_________________________________________"

from math import sqrt, atan, log, tanh

# SI not6ations
mili=10.**(-3)     
micro=10.**(-6)
nano=10.**(-9)
kilo=10.**(3)
Mega=10.**(6)
Giga=10.**(9)

# constants  
pi=4.0*atan(1.)
mu_0=4.0*pi*10**(-7)
c=299792458
eps_0=1.0/(mu_0*c**2)
e=1.60217662*10**(-19)
m_e=9.10938356*10**(-31)
m_eff=10.0 # effective nass of Coopre pairs
m_star=m_eff*m_e # mass of Cooper pair
e_star=2*e
hbar=1.055*10**(-34)
Phi_0=2.*pi*hbar/e_star
#print "e=",e
#print "m_e=",m_e
print "m_eff=",m_eff
#print "c=",c{Fig:Fiory}
#print "mu_0=",mu_0
#print "eps_0=",eps_0
#print "hbar=",hbar
#print "Phi_0=",Phi_0

# geometry of the set-up
a=10.*mili
R1=a/4.0
R2=sqrt(2.0)*R1
R3=sqrt(3.0)*R1
R4=sqrt(4.0)*R1
print "R1=",R1,",     R2=",R2, ",     R3=",R3, ",     R4=",R4 
w=20.*micro
r2=R1+w
r3=R2+w
r4=R3+w
S1=pi*R1**2
S2=pi*(R2**2-r2**2)
S3=pi*(R3**2-r3**2)
S4=pi*(R4**2-r4**2)
d_ins=10.*nano # effective thickness 1/epsilon_relative 100 Angstrom insulator film, AlO_x?
print "d_ins=",d_ins
# capacitors
C1=eps_0*S1/d_ins
C2=eps_0*S2/d_ins
C3=eps_0*S3/d_ins
C4=eps_0*S4/d_ins
C13=C1*C3/(C1+C3)
C24=C2*C4/(C2+C4)
print "C13=",C13,",     C24=",C24


C_d=2.6
# material properties parameterized by lengths 
lambda0=260.*nano # penetration depth at zero temperature
hat_lambda0=lambda0/sqrt(C_d)
xi0=2.6*nano # Ginsburg-Landau extrapolation using B_c2
rho_tot=(eps_0*c**2)*(m_star/e_star)/(lambda0**2)
n_tot=rho_tot/e_star
print "n_tot=",n_tot,",     rho_tot=",rho_tot

# thicness, temperature and values
d_film=100.*nano # superconducting film
T_c=88.0
Delta_T=3.
print "Delta_T=", Delta_T
T=T_c-Delta_T
mod_eps=(T_c-T)/T_c #{Fig:Fiory}
print "mod_eps=",mod_eps
C_s=C_d*mod_eps
print "C_s=",C_s,",     C_d=",C_d

lambdaT=hat_lambda0/sqrt(mod_eps)
xiT=xi0/sqrt(mod_eps)
print "lambdaT=",lambdaT
x=d_film/(2.*lambdaT)
print "x=",x
Z=x/tanh(x)
print "Z=",Z
print "xiT=",xiT
kappa=lambdaT/xiT
print "kappa=",kappa
Bc2=Phi_0/(2.*pi*xiT**2)
Bc=Phi_0/(2.*sqrt(2.)*xiT*lambdaT)
print "Bc2=",Bc2
print "Bc=",Bc
Bc1=Phi_0/(2.*pi*lambdaT**2)*log(10.)
print "Bc1=",Bc1
j_max=(2.*sqrt(2.))/(3.*sqrt(3.))*Bc/(mu_0*lambdaT)
print "j_max=",j_max
B_thin=(m_star/e_star)*log(R2/r2)/((R2**2-r2**2)*C_s*(2.*pi*d_film*rho_tot)**2)
beta=(m_star/e_star)/(2.*rho_tot**2*C_s)
print "!!!     B_thin=",B_thin # according Eq. (9) of version of 4 March; the main result
print "beta=",beta # Eq.(2) of version of 5March

# Frequencies and values of electronic parameters for merasurement
Frequency=250.*Mega
f=250.*kilo
print "f=",f,",   Frequency=", Frequency
omega=2.*pi*f
Omega=2.*pi*Frequency
Y13=1./(Omega*C13) # Imaginary impedance of the drive circuit; small
print "Y13=",Y13
U13a=.5 # amplitude of the basic high-frequency voltage signal
U13b=.5 # amplitude of the modulated voltage signal
E_max=(U13a+U13b)/d_ins # electric field across driving sections
E_breaktrough=1.*Giga
print "E_max=",E_max,",     E_max/E_breaktrough=",E_max/E_breaktrough
Ia=U13a/Y13 # current amplitude of the basic high-frequency signal
Ib=U13b/Y13 # current amplitude of the modulated signal
Q_chgarge=(U13a+U13b)*C13
print "Ia=",Ia,",     Ib=",Ib,",     Q_chgarge=",Q_chgarge
U24=B_thin*Ia*Ib/2. # amplitude of the monochromatic Berboulli voltage

Y24=1./(omega*C24) # impedance of the detector capacitors at low frequency
L24=1./(C24*omega**2) # inductance for resonance detection
print "Y24=",Y24,",     L24=",L24,",     omega*L24=",omega*L24
I24=U24/Y24 # low frequency demodulated current in detector circuit
Ampl=10.**6 # amplification of the low noise pre amplifier; EPO construction
U_amp=U24*Ampl # Bernoully signal to be measured by loc-in
print "I24=",I24

j_2D=(Ia+Ib)/(2*pi*r2) # maximal 2D current density trough the superconducting film
print "j_2D=",j_2D
j=j_2D/d_film # Maximal 3D current density trought the film
print "j/j_max=",j/j_max # (maximal current density)/(depairing current)
print "j=",j
n=n_tot*C_s # volume superfluid density; Cooper pairs per unit volume
v=j/(e_star*n) # maximal drift velocity of Cooper pairs
print "v=",v
v_dep_T=hbar/(m_star*xiT) # maximal depairing velocity at the working temperature
print "v_dep_T=",v_dep_T
v_max_0=hbar/(m_star*xi0) # maximal depairing velocity at zero temperature; order estimate
print "v_max_0=",v_max_0
Q_momentum=m_star*v*xiT/hbar # dimensionless momentum of Cooper pairs in GL units
print "Q_momentum=",Q_momentum
delta_time=5. # lock-in measurements
voltage_noise=1.2*nano # 1.2 nano Volts/sqrt(Hz)
U_noise=voltage_noise*sqrt(delta_time)
signal_noise=U24/U_noise
print "!!!     U24=",U24,",     U_noise=",U_noise,",     signal/noise=",signal_noise
print "U_amp=",U_amp
print "________________________________________"
\end{lstlisting}

Output:
\begin{verbatim}
m_eff= 10.0
R1= 0.0025 ,     R2= 0.00353553390593 ,     R3= 0.00433012701892 ,     R4= 0.005
d_ins= 1e-08
C13= 8.59282330638e-09 ,     C24= 8.50172889227e-09
n_tot= 1.04436343428e+27 ,     rho_tot= 334650935.438
Delta_T= 3.0
mod_eps= 0.0340909090909
C_s= 0.0886363636364 ,     C_d= 2.6
lambdaT= 8.7330788767e-07
x= 0.0572535765518
Z= 1.00109241864
xiT= 1.4081666568e-08
kappa= 62.0173672946
Bc2= 1.66036649352
Bc= 0.0594737755482
Bc1= 0.000994015135589
j_max= 29499277152.6
!!!     B_thin= 3.99429270961e-10
beta= 1.43193264588e-27
f= 250000.0 ,   Frequency= 250000000.0
Y13= 0.0740873807908
E_max= 100000000.0 ,     E_max/E_breaktrough= 0.1
Ia= 6.74878764323 ,     Ib= 6.74878764323 ,     Q_chgarge= 8.59282330638e-09
Y24= 74.8812130373 ,     L24= 4.76708607984e-05 ,     omega*L24= 74.8812130373
I24= 1.21475458408e-10
j_2D= 852.462629601
j/j_max= 0.288977463818
j= 8524626296.01
v= 287.389816813
v_dep_T= 822.449827915
v_max_0= 4454.40932525
Q_momentum= 0.349431426767
!!!     U24= 9.09622967983e-09 ,     U_noise= 2.683281573e-09 ,     signal/noise= 3.38996465051
U_amp= 0.00909622967983
\end{verbatim}

\end{document}